# Representative Pure Risk Estimation by Using Data from Epidemiologic Studies, Surveys, and Registries: Estimating Risks for Minority Subgroups

Lingxiao Wang[1,*], Yan Li[2], Barry I. Graubard[1,^], Hormuzd A. Katki[1,^]

[1]National Cancer Institute, Division of Cancer Epidemiology & Genetics, Biostatistics Branch, Rockville, MD

[2]The Joint Program in Survey Methodology, University of Maryland, College Park, MD.

*Corresponding author lingxiao.wang@nih.gov

^Two authors contributed equally

**SUMMARY**

Representative risk estimation is fundamental to clinical decision-making. However, risks are often estimated from non-representative epidemiologic studies, which usually underrepresent minorities. "Model-based" methods use population registries to improve externally validity of risk estimation but assume hazard ratios (HR) are generalizable from samples to the target finite population. "Pseudoweighting" methods improve representativeness of studies by using an external probability-based survey as the reference, but the resulting estimators can be biased due to propensity model misspecification or inefficient due to variable pseudoweights or small sample sizes of minorities in the cohort and/or survey. We propose a two-step pseudoweighting procedure that poststratifies the event rates among age/race/sex strata in the pseudoweighted cohort to the population rates to produce efficient and robust pure risk estimation (i.e., a cause-specific absolute risk in the absence of competing events). For developing an all-cause mortality risk model representative for the US, our findings suggest that HRs for minorities are not generalizable, and that surveys can have inadequate numbers of events for minorities. Poststratification on event rates is crucial for obtaining reliable risk estimation for minority subgroups.

1  INTRODUCTION

Representative risk estimation plays an increasingly important role in clinical practice and planning public health programs (Adams and Leveson 2012). Currently, much attention is being paid towards improving accuracy of estimates for minority subgroups, especially racial/ethnic minorities, for estimating health disparities and for clinical use of risks for minorities (Vyas et al., 2020). Estimating risk for minorities is challenging because there would naturally be smaller samples sizes for minorities, and worse, most epidemiologic studies further underrepresent minorities (Pinsky et al., 2007). For example, current risk models used to select people for lung cancer screening are substantially less accurate for minorities (Katki et al., 2018).

Hence, estimating risks for minority subgroups would ideally combine data from large datasets, such as research cohorts, including electronic health records, national health surveys, and disease registries. Data sources have strengths/weaknesses. Cohorts usually have all the needed information, but most cohorts consist of volunteers who are generally not representative of the target population, especially for minorities (Pinsky et al., 2007; Fry et al., 2017). Surveys randomly select individuals from the target population, with known selection probabilities to represent the population, but often lack many variables of interest and the sample size of minorities can be small, especially for "rare" outcomes such as cancer mortality. Disease registries are a census of outcomes, typically within age/race/sex strata, and may include mortality, but only summary totals are available.

However, current approaches to estimating risks do not fully use information from all three data sources. One approach is to use individual-level cohort data and summary statistics from a survey or another large database (Chatterjee et al., 2016; Zheng et al., 2022a; 2002b). Most of these methods are "model-based" that aim to build the risk model for outcome prediction, under which the model parameters estimated from any data sources are assumed to be unbiased, and thus can be generalized from the cohort data to the target finite population. However, the generalizability assumption can be violated if the risk model is misspecified or the cohorts are not target-population representative. For example, the Population Attributable Risk (PAR) method (Gail et al., 1989; Chen et al., 2006) is a "model-based" method that assumes hazard ratio (HR) and PAR estimates from the cohort are unbiased and thus generalizable to the target population, and estimates the baseline cumulative hazard by borrowing composite event rates from a disease registry. However, HRs estimated from the cohort can be biased for the population HRs of the same fitted risk model, if the relationships of the risk factors with the outcome variable in the cohort differs from that in the finite population (i.e. "informative" cohort participation (Binder 1992)), or if the fitted risk model is misspecified (Graubard and Korn 1999), and therefore are not generalizable. We found that HRs for minorities estimated from the cohort can substantially differ the sample weighted estimates from the national health survey, or the target population, invalidating the generalizability assumption. Furthermore, PAR in the cohort is unlikely to be generalizable because the PAR depends on the covariate distribution, which is generally not equal in the cohort and the target population.

Another approach is the burgeoning literature on "design-based" methods that improve representativeness of nonprobability samples, such as volunteer-based research cohorts, by assuming an implicit "self-selected participation" mechanism for cohorts and estimate "pseudoweights" (i.e., inverse of the estimated participation rate) for the cohort members from a propensity model by using a probability survey sample as the external reference that represents the target finite population(e.g. Chen et al., 2020; Wang et al, 2021b). However, these methods only use individual data from the cohort and the external reference survey sample, but do not use information from registries. Furthermore, the existing pseudoweighting methods may lead to biased or inefficient risk estimation for subgroups due to two reasons. First, cohort participation is usually assumed to depend on observed covariates only (i.e., noninformative cohort participation conditional on the covariates, Chen et al., 2020; Wang et al., 2021a), as the outcome is often unavailable in the survey. Therefore, these methods only adjust the covariate distribution in the cohort to the weighted survey sample. However, if cohort participation is informative (i.e., associated with the outcome conditional on the covariates) then only adjusting for the covariate distribution without considering the outcome may increase variance of HR estimation without properly reducing bias. Second, although survey samples represent the target population, they may have small sample sizes for minorities, leading to unstable pseudoweighted risk estimation for minorities.

In this paper, we develop a "design-based" two-step pseudoweighting method to estimating pure risk (i.e., a cause-specific absolute risk in the absence of competing events), using all three data sources: individual-level data from a cohort, and a reference survey, and summary statistics from national disease registry, with a special focus on minority subgroups. The first step creates initial pseudoweights for cohort members by fractionally distributing sample weights from survey members to cohort members, according to their kernel smoothed distance in propensity score ("KW.S"). Thus the distribution of covariates in the KW.S pseudoweighted cohort is approximately the same with that in the weighted survey sample which represents the finite population. In the second step, different from traditional poststratification to population totals of covariates in survey research (Valliant et al., 2013), we poststratify on both outcome and covariates to further adjust the KW.S pseudoweights so that the event rate, especially for minority subgroups, in the cohort is approximately the same with that in target population. We call the final pseudoweights "post-KW.S." Taylor series linearization is employed to obtain analytic variance estimates that incorporate all the sources of variability: cohort participation, estimating the propensity model, complex design of the reference survey sample, and poststratification.

The proposed post-KW.S method does not require generalizability assumptions for the naïve cohort estimators because the post-KW.S pseudoweights improve target-population representativeness of the cohort for all considered outcome and covariates, including minority subgroups. Moreover, the post-KW.S method, by poststratifying to subgroup event rates reduces reliance on both noninformative cohort participation, and on the survey with inadequate sample size for age/race/sex subgroups. We simulate to investigate the impact of violations of generalizability assumptions and noninformative cohort participation. We applied the proposed method to develop a representative risk model of all-cause mortality for the general US population, focusing on minority subgroups, by using data from the NIH-AARP epidemiologic cohort, National Health Interview Survey, and poststratifying to CDC registry mortality rates by age/race/sex cells.

## 2  BASIC SURVIVAL MODEL, POPULATIONS, AND SAMPLING

Pure risk (Shin et al., 2020; Zheng et al., 2022b) is the probability of the event of interest occurring by time $t$: $r(t) = P(0 < T \leq t) = 1 - S(t)$, where $T$ is the time to the event. The time-on-study is used as the time scale, with the baseline of the study being treated as $t = 0$. The survival function $S(t) = \exp\{-\Lambda(t)\}$, with $\Lambda(t)$ being the cumulative hazard function for $T$ that is absolutely continuous and differentiable, and $\Lambda(t) = \int_0^t \lambda(\tau) d\tau$ where $\lambda(t) = \lim_{\epsilon \to 0^+} P(t \leq T < t + \epsilon \mid T \geq t) / \epsilon$ is the hazard at time $t$. Under a Cox regression model (Cox 1972), the hazard function is $\lambda(t \mid \mathbf{z}) = \lambda_0(t) \exp(\boldsymbol{\beta}^T \mathbf{z})$, where $\mathbf{z}$ is a vector of non-time dependent covariates, $\lambda_0(t)$ is the baseline hazard, and $\boldsymbol{\beta}$ is a vector of the log hazard-ratios.

## 2.1 Finite Population

The target finite population ($FP$) consists of $M$ individuals indexed by $i \in \{1, \cdots, M\}$. Each individual $i$ has a vector of non-time dependent covariates $\mathbf{z}_i$ (including the subgroup indicators and other covariates), a potential event time $T_i$ or censoring time $C_i$, and the observed time $X_i = \min(T_i, C_i)$. We assume that the potential event time $T$ and the censoring time $C$ are independent conditional on the covariates $\mathbf{Z}$. Denote the event counting process $N_i(t)$, ($N_i(t) = 1$ if $T_i < t$, and $T_i \leq C_i$; and $N_i(t) = 0$ otherwise), and the at-risk process, $Y_i(t)$, ($Y_i(t) = 1$ if $X_i \geq t$, and $Y_i(t) = 0$ otherwise). The time metric is follow-up time with a maximum of $t_0$. The event status for each individual $i$ during a given follow-up period $[0, t_0]$ is $D_i$, ($D_i = 1$ if $T_i = \min(T_i, C_i, t_0)$, and $D_i = 0$ otherwise). The event and non-event populations during the follow-up period are $FP_1 = \{i \mid D_i = 1, i \in FP\}$, with size $M_1$, $FP_0 = \{i \mid D_i = 0, i \in FP\}$, with size $M_0$, respectively. We are interested in estimating the pure risk for subgroups controlling for other covariates at time $t$ under the Cox regression model (Cox 1972) for an individual in the $FP$ subgroups of interest with covariate $\mathbf{z}^\dagger$

$$r^{FP}(t, \mathbf{z}^\dagger, \boldsymbol{\beta}_{FP}, \Lambda_0^{FP}) = 1 - \exp\{-\Lambda_0^{FP}(t) \cdot \exp(\boldsymbol{\beta}_{FP}^T \mathbf{z}^\dagger)\}, \tag{2.1}$$

where $\boldsymbol{\beta}_{FP}$ is a vector of $FP$ log HRs obtained from a Cox regression model, and $\Lambda_0^{FP}(t)$ is the baseline cumulative hazard at time $t$.

## 2.2 Cohort and Survey Sample Selection

Let $s_c \subset FP$ denote a cohort with $n_c$ individuals. Similarly to (Chen et al., 2020), we define a random indicator variable $\delta^{(c)}$ for individuals in $FP$ ($\delta_i^{(c)} = 1$ if $i \in s_c$, and $\delta_i^{(c)} = 0$ if $i \in FP - s_c$) that specifies which individuals in $FP$ participate in $s_c$. $FP$ and $s_c$ are also used to denote sets of indices for the $FP$ and the cohort, respectively. We assume that the cohort members have an "implicit" unobserved participation mechanism resulting in a cohort participation rate $\pi_i^{(c)} \equiv \Pr(i \in s_c \mid FP) = E_c\{\delta_i^{(c)}\}$, where the expectation $E_c$ is with respect to the unknown cohort sample participation process from $FP$. The corresponding cohort implicit (unknown) sample weight is $w_i^{(c)} = 1/\pi_i^{(c)}$ for $i \in s_c$. We require standard assumptions of positive cohort participation $\pi_i^{(c)} > 0$.

In addition, we assume that there is a reference survey sample $s_s$ that denotes the subset of indices for $n_s$ individuals in the survey sample that are randomly sampled from the $FP$. The sample inclusion indicator, inclusion probability, and corresponding sample weights are defined by $\delta_i^{(s)}$ ($\delta_i^{(s)} = 1$ if $i \in s_s$; $\delta_i^{(s)} = 0$ if $i \in FP - s_s$), $\pi_i^{(s)} = E_s\{\delta_i^{(s)}\}$, and $w_i^{(s)} = 1/\pi_i^{(s)}$, respectively, where $E_s$ is the expectation over all possible random samples according to the survey sample design. For the survey sample, $\{\pi_i^{(s)}, i \in s_s\}$ are known by the sample design and are assumed to include adjustments for nonresponse and calibration for $FP$ quantities, e.g., population totals by age, race/ethnicity and sex categories. We assume that all $FP$ members have positive survey sample selection probability, i.e., $\pi_i^{(s)} > 0$ for $i \in FP$.

## 3  ESTIMATING THE RISK MODEL

### 3.1  Two-Step Pseudoweighting Method

Pseudoweights are not formal sampling weights $\{w_i^{(c)}, i \in s_c\}$, but can improve representativeness of the cohort, $s_c$ (Chen et al., 2020; Wang et al., 2020; 2021a, b). We propose a two-step pseudoweighting method. The first step computes pseudoweights for the cohort, $\{\widehat{w}_i, i \in s_c\}$, using a scaled kernel-weighting (KW.S) method based on covariates $\mathbf{z}$ (including the indicator for minorities) that are available in both $s_c$ and $s_s$. The second step is poststratification weighting adjustment of the KW.S pseudoweighted $s_c$ to the $FP$ by the event rate in poststrata defined by covariates $\mathbf{z}_0$, which is usually a subset of $\mathbf{z}$ including race/ethnicity and other demographic variables such as age and sex, that are available in the registry.

*3.1.1 First Step: Compute Pseudoweights for Cohort Using Kernel-Weighting Method*

The first step aims to pseudoweight the cohort so that the covariate distribution in the pseudoweighted cohort approximates to that in the finite population. We fit a logistic propensity model in the combined naïve cohort and **weighted** survey sample

$$\log\left\{\frac{\Pr(R_i = 1)}{1 - \Pr(R_i = 1)}\right\} = \gamma_0 + \boldsymbol{\gamma}^T \mathbf{z}_i, \text{ for } i \in s_c \cup s_s, \tag{3.1}$$

where $R_i$ indicates being in the cohort or survey ($R_i = 1$ if $i \in s_c$, 0 if $i \in s_s$), and $(\gamma_0, \boldsymbol{\gamma}^T)^T$ is a vector of unknown parameters, which can be estimated from the scaled weighted estimating equation

$$\psi(\boldsymbol{\gamma}) = \sum_{i \in s_c} w_i (1 - p_i)(1, \mathbf{z}_i^T)^T - a \sum_{i \in s_s} w_i p_i (1, \mathbf{z}_i^T)^T = \mathbf{0}, \tag{3.2}$$

Where $w_i = 1$ for $i \in s_c$, $w_i = w_i^{(s)}$ for $i \in s_s$), $p_i = \text{expit}(\gamma_0 + \boldsymbol{\gamma}^T \mathbf{z}_i)$ is the propensity score, and $a = n_s/M$ is a scaling factor that reduces the variation of the weights in the combined sample and thereby improves the efficiency of the resulting estimator, denoted by $(\hat{\gamma}_0, \hat{\boldsymbol{\gamma}}^T)^T$ (Li et al., 2011; Wang et al., 2021a). Note that $E(\hat{\gamma}_0) \doteq \gamma_0 + \log(a)$, and $E(\hat{\boldsymbol{\gamma}}) \doteq \boldsymbol{\gamma}$, where $(\gamma_0, \boldsymbol{\gamma}^T)^T$ is the true values of parameters in equation (3.1). That is, rescaling the survey weights by factor $a$ for the propensity modelling only affects the intercept of the coefficients, $\gamma_0$, which can be corrected by the offset of $\log(a)$. (Wang et al., 2021a).

To construct pseudoweights $\{\widehat{w}_i, i \in s_c\}$, each survey member first assigns a fraction of the sample weight to all cohort members proportionally to the kernel distance, measured by logit of propensity scores, so that most of its sample weight is assigned to those cohort members with similar propensity scores. The assigned portions of the weight from each survey member are then summed up to form pseudoweights for cohort members. The details are as follows:

$$\widehat{w}_i = \sum_{j \in s_s} \left\{ \frac{K\left\{(q_i^{(c)} - q_j^{(s)})/h\right\}}{\sum_{k \in s_c} K\left\{(q_k^{(c)} - q_j^{(s)})/h\right\}} \cdot w_j^{(s)} \right\}, \tag{3.3}$$

where $q_i^{(c)} = \boldsymbol{\gamma}^T \boldsymbol{z}_i$ and $q_j^{(s)} = \boldsymbol{\gamma}^T \boldsymbol{z}_j$ for $i \in s_c$ and $j \in s_s$ respectively, with $\boldsymbol{\gamma}$ estimated from equation (3.2), $K(\cdot)$ is the kernel function where we use standard normal density function, and $h = 0.9 \min(\hat{\sigma}, IQR/1.34) \cdot n_c^{-1/5}$ is the bandwidth selected by Silverman's rule of thumb method (Silverman 1986), with $\hat{\sigma}$ and $IQR$ being the standard deviation and interquartile range of $\left\{q_i^{(c)}; i \in s_c\right\}$ respectively.

### 3.1.2 Second step: Poststratification Adjusting the Pseudoweighted Cohort to the Population

The traditional poststratification using population totals of subgroups defined by the covariates only has been shown to obtain limited bias reduction or efficiency improvement for association estimation (Lumley et al., 2011). Hence, at the second step, we poststratify the KW.S pseudoweighted cohort to the population with the poststrata defined by both event status $D$ and the subgroup indicator $\boldsymbol{z_0}$ (usually demographic variables of race/age/sex) that are available in the $FP$ and the event registry, so that the event rate in each subgroup (or poststratum) in the pseudoweighted $s_c$ is consistent with that in the $FP$. We use $s_{c,1} = \{i, D_i = 1, i \in s_c\}$), and $s_{c,0} = \{i, D_i = 0, i \in s_c\}$ to denote sets of events and non-events (including censored individuals) in the cohort respectively. We then form $G$ poststrata by $\boldsymbol{z_0}$ in $s_{c,1}$, $s_{c,0}$, $FP_1$ and $FP_0$. The poststratification weight factor for $i \in s_c$ is

$$f_i = \frac{M_{d,g}}{\sum_{j \in s_{c,d}^{(g)}} \widehat{w}_j}, \quad \text{for } i \in s_{c,d}^{(g)} \tag{3.4}$$

where $g = 1, \cdots G$ is the $g^{\text{th}}$ poststrata, $M_{d,g}$ is the size for $g^{\text{th}}$ poststrata in $FP_d$, which are obtained from the event registry, and $s_{c,d}^{(g)}$ is the $g^{\text{th}}$ poststrata in $s_{c,d}$, for $d = 0, 1$. The final weights from post-KW.S method for the cohort from the two-step approach are $\{\widetilde{w}_i = f_i \cdot \widehat{w}_i, i \in s_c\}$.

## 3.2 Cohort Estimates of the Hazard Ratio and Pure Risk

The log hazard-ratios $\boldsymbol{\beta}$ are estimated from the post-KW.S estimating equation in the cohort

$$\widetilde{U}(\boldsymbol{\beta}) = \sum_{i \in s_c} \int_0^\infty \widetilde{w}_i \{\mathbf{z}_i - \hat{S}^{(0)}(\tau, \boldsymbol{\beta})^{-1} \hat{S}^{(1)}(\tau, \boldsymbol{\beta})\} dN_i(\tau) = \mathbf{0}, \qquad (3.5)$$

where $\widetilde{w}_i$ is the post-KW.S for cohort unit $i$ proposed in Section 3.2, $\hat{S}^{(u)}(\tau, \boldsymbol{\beta}) = M^{-1} \sum_{j \in s_c} \widetilde{w}_j Y_j(\tau) \exp(\boldsymbol{\beta}^T \mathbf{z}_j) \cdot \mathbf{z}_j^{\otimes u}$, with $\mathbf{z}_j^{\otimes 0} = 1$, $\mathbf{z}_j^{\otimes 1} = \mathbf{z}_j$, and $dN_i(\tau) = N_i(\tau) - N_i(\tau-)$ is the increment of $N_i$ at time $\tau$, with $N_i(\tau-) = \lim_{\epsilon \to 0^+} N_i(\tau - \epsilon)$. Denote the solution of (3.5) as $\widehat{\boldsymbol{\beta}}_{\widetilde{w}}$.

Then we have the following result, proved in Web Appendix A.2:

**Theorem** *Under conditions* **C1-C8** *in the Web Appendix A.1 and assuming the logistic regression model* (3.1) *is correctly specified for the propensity scores, $\widehat{\boldsymbol{\beta}}_{\widetilde{w}}$ is a consistent estimator of the finite population parameter $\boldsymbol{\beta}_{FP}$, i.e., $\widehat{\boldsymbol{\beta}}_{\widetilde{w}} = \boldsymbol{\beta}_{FP} + O_p(n_c^{-1/2})$.*

The post-KW.S Breslow baseline cumulative hazard estimator (Breslow, 1972) is

$$\widehat{\Lambda}_0^B(t, \widehat{\boldsymbol{\beta}}_{\widetilde{w}}) = \int_0^t \frac{d\widehat{N}(\tau)}{\hat{S}^{(0)}(\tau, \widehat{\boldsymbol{\beta}}_{\widetilde{w}})}, \qquad (3.6)$$

where $d\widehat{N}(\tau) = M^{-1} \sum_{i \in s_c} \widetilde{w}_i dN_i(\tau)$, and $\hat{S}^{(0)}(\tau, \widehat{\boldsymbol{\beta}}_{\widetilde{w}})$ defined in equation (3.5) with $\boldsymbol{\beta} = \widehat{\boldsymbol{\beta}}_{\widetilde{w}}$.

The post-KW.S Breslow estimator can be inefficient due to the pseudoweighting and limited sample size of cohort or the reference survey sample. Analogous to Gail (1989), but in the absence of competing events, we propose a post-KW.S PAR estimator of the baseline cumulative hazard by including composite event rates from a registry as follows

$$\widehat{\Lambda}_0^P(t, \widehat{\boldsymbol{\beta}}_{\widetilde{w}}) = \int_0^t \{1 - \widehat{AR}(\tau, \widehat{\boldsymbol{\beta}}_{\widetilde{w}})\} \cdot \lambda^*(\tau) \, d\tau, \tag{3.7}$$

where $\widehat{AR}(\tau, \widehat{\boldsymbol{\beta}}_{\widetilde{w}}) = 1 - \widehat{S}^{(0)}(\tau, \widehat{\boldsymbol{\beta}}_{\widetilde{w}})^{-1} \widehat{S}^{(0)}(\tau)$ is the post-KW.S estimator of the "population attributable risk", with $\widehat{S}^{(0)}(\tau) = M^{-1} \sum_{i \in S_c} \widetilde{w}_i Y_i(\tau)$, and $\lambda^*(\tau) = dN(\tau)/S^{(0)}(\tau)$ is the composite rate of the event in the $FP$ at time $\tau$, with $dN(\tau) = M^{-1} \sum_{i \in FP} dN_i(\tau)$, and $S^{(0)}(\tau) = M^{-1} \sum_{i \in FP} Y_i(\tau)$ obtained from the event registry. For all-cause mortality, the event set $\sum_{i \in FP} dN_i(\tau)$ is number of people dying at time $\tau$, and the risk set $\sum_{i \in FP} Y_i(\tau)$ is the number of alive individuals in the population before time $\tau$. We have the following result, proved in Web Appendix A.3:

**Corollary** *Under* **C9** *and* **C10** *in Web Appendix A.1 and conditions required by* **Theorem**, $\widehat{\Lambda}_0^B(t, \widehat{\boldsymbol{\beta}}_{\widetilde{w}})$ *and* $\widehat{\Lambda}_0^P(t, \widehat{\boldsymbol{\beta}}_{\widetilde{w}})$ *are consistent estimators of* $\Lambda_0^{FP}(t, \boldsymbol{\beta})$, *i.e.,* $\widehat{\Lambda}_0^B(t, \widehat{\boldsymbol{\beta}}_{\widetilde{w}}) = \Lambda_0^{FP}(t, \boldsymbol{\beta}_{FP}) + O_p(n_c^{-1/2})$; $\widehat{\Lambda}_0^P(t, \widehat{\boldsymbol{\beta}}_{\widetilde{w}}) = \Lambda_0^{FP}(t, \boldsymbol{\beta}_{FP}) + O_p(n_c^{-1/2})$ *uniformly at time t.*

Finally, the pure risk estimator at time $t$ for a subgroup of interest, defined by covariate $\mathbf{z}^\dagger$ is

$$\hat{r}(t, \mathbf{z}^\dagger, \widehat{\boldsymbol{\beta}}_{\widetilde{w}}, \widehat{\Lambda}_0) = 1 - \exp\{-\widehat{\Lambda}_0(t) \cdot \widehat{rr}\}, \tag{3.8}$$

by using $\widehat{\Lambda}_0 = \widehat{\Lambda}_0^B$ or $\widehat{\Lambda}_0^P$, where $\widehat{rr} = \exp(\widehat{\boldsymbol{\beta}}_{\widetilde{w}}^T \cdot \mathbf{z}^\dagger)$ is the post-KW.S estimator of the relative risk (i.e. hazard ratio). Since $\hat{r}$ is a continuous function of $\widehat{\Lambda}_0(t)$, and $\widehat{\boldsymbol{\beta}}_{\widetilde{w}}$, which are consistent, $\hat{r}$ is also a consistent estimator of $r^{FP}(t, \mathbf{z}^\dagger, \boldsymbol{\beta}_{FP}, \Lambda_0^{FP})$ in formula (2.1). We note that the consistency of $\hat{r}$ holds for any $FP$ risk model specified in formula (2.1).

### 3.3 Variance Estimation Using Taylor Deviates Accounting for All Sources of Variability

The finite population variance of an estimator, $\hat{\theta}$, can be approximated by the variance of the sum of the influence operators, $\Delta_i(\hat{\theta})$, i.e., $var(\hat{\theta}) \approx var\{\sum_{i=1}^n \Delta_i(\hat{\theta})\}$ for simple random samples (van der Vaart, 1998, Chapter 20). It has been extended to estimating variances under complex sample designs by using Taylor deviates $\Delta_i(\hat{\theta}) = w_i \cdot \partial \hat{\theta}/\partial w_i$ as the weighted influence operators (Graubard & Fears, 2005), with $w_i$ being the sample weights for sample unit $i$. We adapt this method to estimate variance of $\hat{r}$, accounting for all sources of variation due to cohort participation, survey sampling, propensity estimation, and poststratification.

Consider the general case where the reference survey sample $s_s$ is selected under a stratified multistage cluster sampling with $H$ sampling strata and $u_h$ primary sample units (PSUs) in stratum $h$, $h = 1, \cdots, H$. We combine the cohort $s_c$ with the $s_s$ by treating $s_c$ as stratum, $H + 1$ independent from $s_s$, in the combined sample $s_c \cup s_s$ (under condition **C12** in Web Appendix A, assuming the cohort and the survey sample selection are uncorrelated). In $s_c \cup s_s$. The Taylor linearization estimator, $var(\hat{r})$, is obtained by

$$var(\hat{r}) = var\left\{\sum_{i \in s_c \cup s_s} \Delta_i(\hat{r})\right\} = \sum_{h=1}^{H+1} \frac{u_h}{u_h - 1} \sum_{i=1}^{u_h} (v_{hi} - \bar{v}_h)^2, \tag{3.9}$$

where $v_{hi} = \sum_{j=1}^{u_{hi}} \Delta_{hij}(\hat{r})$ is the sample total of the Taylor deviates for cluster $hi$, and $\bar{v}_h = (u_h)^{-1} \sum_{i=1}^{u_h} v_{hi}$. Independent cohort individuals are treated as clusters, i.e., $u_{H+1} = n_c$, although (3.9) this also allows clustering or stratification for $s_c$.

In order to account for randomness due to pseudoweight assignment and poststratification, we write post-KW.S weights as functions of the original weights $\{w_i, i \in s_c \cup s_s\}$ which are used for propensity estimation in equation (3.2) and obtain $\Delta_i(\hat{r})$ as follows

$$\Delta_i(\hat{r}) = w_i \frac{\partial \hat{r}}{\partial w_i} = -\exp\{-\widehat{\Lambda}_0(t, \widehat{\boldsymbol{\beta}}_{\widetilde{w}}) \cdot \widehat{rr}\} \cdot \widehat{rr}[\mathbf{z}^T \Delta_i(\widehat{\boldsymbol{\beta}}_{\widetilde{w}})\widehat{\Lambda}_0(t, \widehat{\boldsymbol{\beta}}_{\widetilde{w}}) + \Delta_i\{\widehat{\Lambda}_0(t, \widehat{\boldsymbol{\beta}}_{\widetilde{w}})\}], \tag{3.10}$$

for $i \in s_c \cup s_s$, where $\hat{r}$ can be $\hat{r}^B$ or $\hat{r}^P$, depending on the choice of baseline cumulative hazard estimator $\widehat{\Lambda}_0(t, \widehat{\boldsymbol{\beta}}_{\widetilde{w}}) = \widehat{\Lambda}_0^B(t, \widehat{\boldsymbol{\beta}}_{\widetilde{w}})$ or $\widehat{\Lambda}_0^P(t, \widehat{\boldsymbol{\beta}}_{\widetilde{w}})$ respectively, $\Delta_i(\widehat{\boldsymbol{\beta}}_{\widetilde{w}})$ and $\Delta_i\{\widehat{\Lambda}_0(t, \widehat{\boldsymbol{\beta}}_{\widetilde{w}})\}$ are the Taylor deviates for the log-hazard ratios and cumulative baseline hazards, respectively (Web Appendix B).

# 4 SIMULATION STUDIES

## 4.1 Finite population, registry, cohort and survey sample generation

We generated a finite population $FP$ of size $M = 200,000$. Without loss of generality, we generated three continuous covariates $z_1 \sim N(0, 4)$, $z_2 \sim N(0, 1.5)$, and $z_3 \sim N(0, 1)$. actual time to event variable was generated by $T \sim Weibull\{\theta, \alpha = 1\}$ with $\theta = \exp(\beta_0 + \boldsymbol{\beta}^T \boldsymbol{z})$, where $\boldsymbol{\beta} = (\beta_1, \beta_2, \beta_3)^T$, and $\boldsymbol{z} = (z_1, z_2, z_3)^T$. The hazard function is $\lambda(t; \boldsymbol{z}) = \lambda_0(t) \exp(\boldsymbol{\beta}^T \boldsymbol{z})$ is time-invariant, with the baseline hazard $\lambda_0(t) = \exp(\beta_0)$. We set $\beta_0 = \log\{-(\log 0.95)/15\}$, and $\boldsymbol{\beta} = (0.25, 0.4, 0.15)$. We set administrative censoring at 15 years after start of the study and $C_1 = 15 - T_0$ was the time from entry to administrative censoring, where $T_0 \sim U(0, 1)$ is the random starting time. In simulations, the event of interest is not limited to all-cause mortality and can be any specific diseases. Hence, we also considered censoring from death due to other causes, $C_2 \sim Weibull\{\theta = -(\log 0.9)/15, \alpha = 1\}$, which is independent from the actual event $T$. The observed time was defined as $X = \min(T, C_1, C_2)$, and the indicator for event of interest was $D = I\{T \leq \min(C_1, C_2)\}$. In the generated $FP$, the proportion of the event of interest (i.e., $D = 1$) and the proportion of death due to other causes (i.e., $X = C_2$) were 8.12% and 9.29% respectively, during the follow-up. We are particularly interested in estimating HR of $z_2$, i.e., $\exp(\beta_2)$, controlling for $(z_1, z_3)$, and the pure risks $\boldsymbol{r} = (r_{low}, r_{med}, r_{high})$ for three types of individuals from either low (25$^{th}$ percentile), medium (median), and high (75$^{th}$ percentile) values of $z_2$.

The event registry had the event set in the $FP$, i.e., $FP_1 = \{i \mid D_i = 1, i \in FP\}$, with size $M_1$, and the non-event set was $FP_0 = FP - FP_1$. To obtain more robust and efficient estimators of $\beta_2$ and $\boldsymbol{r}$, poststrata should be formed by $D$ and variables highly correlated with $z_2$. We assumed that only the categorized covariate $z_2$, denoted by $z_2^*$ ($z_{2,i}^* = 1$ if $z_{2,i} < 0$, and $z_{2,i}^* = 2$ otherwise), was available in the registry and $FP$ for poststratification. A volunteer cohort, $s_c$, of $n_c = 5{,}000$ was randomly selected using Probability Proportional to Size (PPS) sampling with the measure of size $sz^{(c)} = \exp(\gamma_1 z_1 + \gamma_2 z_2 + \gamma_d D + \gamma_{2,d} z_2 \cdot D)$, i.e., the cohort participation rate for $FP$ unit $i$ is proportional to the measure of size, $\pi_i^{(c)} \propto sz_i^{(c)}$. Although we generated random samples of cohorts, cohort recruitment in reality is usually volunteer based, with unobserved cohort participation rates. Hence, in the analysis, we masked the PPS design and treated the cohort as a simple random sample for naïve risk estimation, as is typical in cohort analyses. Then we apply the proposed post-KW.S method to estimate $\{w_i^{(c)} = \pi_i^{(c)}, i \in s_c\}$ for risk estimation. We varied values of $\gamma_d$, and $\gamma_{2,d}$ under three scenarios (Table 1, Web Appendix C) to investigate how the "implicit" cohort participation mechanism influenced cohort estimates of $\beta_2$ and $r$, and how post-KW.S can help reduce bias and improve efficiency. In the naïve (i.e., unweighted) cohort, the averaged proportion of observed event of interest was 11%-16%, varied due to the different cohort selection in the three scenarios. The proportion of observed competing event $C_2$ in the naïve cohort was ~9.3% in all three scenarios. A survey of size $n_s = 3{,}000$ was randomly selected using PPS sampling with selection probability $\pi_i^{(s)} \propto \exp(0.07 z_1 + 0.1 z_2)$ in all three scenarios. The survey sample design was explicit and considered in the analysis. We assumed the outcome variables $T$ and $D$ were **not** available in the survey sample.

### 4.2 Simulation results

We compared the performance of the naïve, KW.S, and post-KW.S methods in estimating $\boldsymbol{\beta}$, $\Lambda_0$, and $\boldsymbol{r}$ at $t = 1$, with the results shown in Table 1, Web Figures 1-3, and Table 2 respectively. When calculating the bias of sample estimates of $\boldsymbol{\beta}$, $\Lambda_0$, and $\boldsymbol{r}$, we use the $FP$ quantities $\boldsymbol{\beta}^{FP}$, $\Lambda_0^{FP}$, and $\boldsymbol{r}^{FP}$ in equation (2.1) as the true values. In scenarios 2 and 3 where the "implicit" cohort participation depended on the disease status, we presented results under both the correctly specified and a misspecified propensity model that missed the outcome variable.

As "implicit" cohort participation depended on the main effect of $z_1$, and did not depend on $z_3$, the naïve cohort estimates of $\beta_1$ and $\beta_3$ were approximately unbiased in all scenarios. The bias of naïve cohort estimates of $\beta_2$, $\Lambda_0$, and $r$ (denoted by $\hat{\beta}_2^{\text{naive}}$, $\hat{\Lambda}_0^{B,\text{naive}}$, $\hat{\Lambda}_0^{P,\text{naive}}$, and $\hat{r}^{\text{naive}}$ respectively) varied in the three scenarios due to the different "implicit" cohort participation mechanisms. In scenario 1, the cohort participation was noninformative (i.e., associated with $z$ only). Hence, $\hat{\beta}_2^{\text{naive}}$ and $\hat{\Lambda}_0^{B,\text{naive}}$, were approximately unbiased, and generalizable from $s_c$ to the $FP$, leading to unbiased $\hat{r}^{\text{naive}}$. In scenario 2, the "implicit" cohort participation was associated with both $z_2$ and disease status $D$, marginally. The $\hat{\beta}_2^{\text{naive}}$ was slightly biased, but the bias in $\hat{\Lambda}_0^{B,\text{naive}}$ increased by ~30% over follow-up, since the event rate in the $s_c$ was 36.5% higher than that in the $FP$.. The $\hat{r}^{\text{naive}}$ overestimated $r_{\text{low}}$, $r_{\text{med}}$, and $r_{\text{high}}$ by 36.9%, 29.5%, and 22.7% respectively. In scenario 3, both $\hat{\beta}_2^{\text{naive}}$ and $\hat{\Lambda}_0^{B,\text{naive}}$ were biased because the "implicit" cohort participation was informative (i.e., associated with the joint distribution of $z_2$ and $D$). As a result, $\hat{r}^{\text{naive}}$ had the worst bias among all the scenarios.

The naïve PAR cohort estimator of $\Lambda_0$, denoted by $\hat{\Lambda}_0^{P,\text{naive}}$, was biased by -35% in all scenarios and therefore were not generalizable from $s_c$ to the $FP$. This is because the naïve PAR itself was biased due to different covariate distributions in $s_c$ from the $FP$. However, $\hat{\Lambda}_0^{P,\text{naive}}$, by incorporating composite event rate from the $FP$, was more efficient than $\hat{\Lambda}_0^{B,\text{naive}}$ in all scenarios and yielded smaller mean-squared error (MSE) in scenarios 2 and 3. Furthermore, unlike $\hat{\Lambda}_0^{B,\text{naive}}$, $\hat{\Lambda}_0^{P,\text{naive}}$ was not sensitive to the biased event rate in the cohort. We varied the event rate in the cohort in more scenarios. The relative bias of $\hat{\Lambda}_0^{B,\text{naive}}$ jumped from 0% to 60% as the event rate increased from 11.5% to 18.5% in the cohort, while the relative bias of $\hat{\Lambda}_0^{P,\text{naive}}$ remained -35% (Web Figure 4).

Post-KW.S yielded approximately unbiased and most efficient estimates for $\beta_2$, $\Lambda_0$ and $r$ in all scenarios under the correct propensity model. The post-KW.S PAR method sharply reduced the variance of $\widehat{\Lambda}_0^{B,\text{naive}}$, leading to the largest reduction in MSE of $\hat{r}^{\text{naive}}$ among all considered methods. In Scenario 3 where the "implicit" cohort participation was informative, the KW.S estimator did not reduce any bias of $\hat{\beta}_2^{\text{naive}}$ and had a large variance under the misspecified model that missed the outcome $D$ which was assumed to be unavailable in the survey. On the contrary, the post-KW.S method, by poststratifying to the subgroup event rates in the $FP$, greatly reduced the bias of $\hat{\beta}_2^{\text{naive}}$ and $\hat{r}^{\text{naive}}$. The TL variance estimates were close to the empirical variances. We also varied the correlation among covariates up to 0.6. The pattern of results was similar. (Web Tables 1-3).

## 5 DEVELOPING A NATIONALLY REPRESENTATIVE ALL-CAUSE MORTALITY PURE RISK MODEL FOR MINORITIES

We develop an risk model for all-cause mortality using the NIH-AARP Diet and Health Study, which recruited $n_c = 567{,}169$ AARP members among 8 states from 1995-1996, ages 50 to 71 years (NIH-AARP, 2006). We are especially interested in estimating HRs and pure risks for minorities. For the reference survey, we used the NHIS, a cross-sectional household interview survey of the civilian noninstitutionalized US population. To make the two samples as contemporaneous as possible, we chose the 1997 NHIS respondents aged 50 to 71 years ($n_s = 9{,}306$ participants). The 1997 NHIS has a multistage stratified cluster sample design with 339 strata with each consisting of two sampled PSUs. In this particular example, both NIH-AARP and NHIS were linked to National Death Index (NDI) for ten-year follow-up mortality information (NCHS 2009; NCHS 2013), enabling us to examine how much bias in the NIH-AARP estimates can be corrected (assuming the NHIS is the "gold standard") by the proposed methods, and how much variance in NHIS estimates can be reduced. Finally, we used CDC mortality rates 1999-2009 (CDC, 2021) as the population composite mortality rates and for poststratification. The NIH-AARP cohort has substantially lower mortality than CDC rates in all age groups ("healthy volunteer effect"). For minorities, mortality rates were underestimated for Non-Hispanic Blacks but overestimated for non-Hispanic others in the NIH-AARP. The NHIS estimates appeared unbiased relative to CDC rates, but with considerable variance, especially for non-Hispanic Blacks (282 deaths), Hispanics (143 deaths) and non-Hispanic other (40 deaths), due to small sample sizes for mortality (Figure 1, Web table 4).

The all-cause mortality risk model included race/ethnicity, along with age, sex, smoking, physical activity, BMI, and self-reported health status, a highly predictive factor for mortality (Gill 2012). We fit the propensity models to the combined NIH-AARP and weighted NHIS 1997 that included covariates with different distributions in the naïve NIH-AARP and NHIS and are also associated with all-cause mortality (Li et al., 2022). Thus, we included all mortality model covariates, as well as education and marital status (which were excluded from the mortality model as being clinically unavailable). The propensity model parameters (Web Table 5) show that, in AARP, the young, women, minorities, and people with "poor" health status are greatly underrepresented. KW.S nearly equalized the covariate distribution in the AARP to the weighted NHIS (Web Table 6). However, KW.S underestimated the population mortality rate. The joint distribution of covariates and mortality status in the KW.S weighted AARP was still different from that in the CDC registry (Web Table 4), suggesting lack-of-fit of the propensity model. Post-KW.S, by poststratifying the KW.S weighted AARP to the CDC registry, increased the mortality rate and corrected the distribution of demographic variables by mortality status, especially for minorities (Web Table 4). The all-cause risk model was fitted using the above covariates from the baseline NIH-AARP cohort.

Pseudoweighting and poststratification were most important for HRs for minorities (Table 3). For non-Hispanic Blacks, NIH-AARP had an implausible HR=1.03 (Xu et al., 2022), but was HR=1.26 using post-KW.S. For non-Hispanic others (who are mostly Asian-American), NIH-AARP had an implausible HR=1.01, and even the NHIS had an implausible HR=0.95 (Xu et al., 2022), probability due to few cases of mortality among non-Hispanic others ($n$=40 in Web table 4), but the HRs post-KW.S HR was 0.65, which is substantially smaller, as expected. Finally, the post-KW.S minority HRs had narrower confidence intervals than the KW.S, especially for Non-Hispanic Black and Non-Hispanic Other, and substantially narrower than NHIS.

Table 4 shows the ratios of expected pure risk of 10-year mortality (calculated by the (weighted) sample mean of estimated pure risks) to observed 10-year mortality rate in the CDC registry. The NHIS overestimated risks for Hispanics and non-Hispanic others. The naïve AARP-Breslow substantially underestimated the risks in nearly all poststrata, but pseudoweighting-alone (KW.S) provided well-calibrated estimates, but notably, underestimated the risk for non-Hispanic blacks and substantially overestimated the risk for non-Hispanic-others. Post-KW.S performed accurately for all poststrata and generally yielded the closest average risk to CDC mortality among all methods, including the NHIS, especially for minority groups.

Table 5 shows example of pure risk estimates for three minorities with low, medium, and high risks, with the TL standard errors. NHIS-PAR estimates had substantial variability due to the small sample size. The naïve AARP-Breslow estimates were substantially different from the NHIS estimates for all three individuals due to the biased HRs and baseline cumulative hazards (Web Figure 5). The KW.S-PAR estimates were very close to AARP-PAR estimates, but with inflated variance due to the differential pseudoweights. In contrast, the post-KW.S-PAR estimates were closer to the NHIS-PAR estimates, and were more efficient than the KW.S estimates, especially for the medium and high risks. Notice that the Post-KW.S-PAR estimate of pure risk is smaller than the NHIS-PAR estimate for Non-Hispanic, because NHIS overestimated the HR and the pure risk for Non-Hispanic (Table 3, Table 4). The pattern of pure risk estimates for racial/ethnic minorities in the NIH-AARP cohort using the HR's and baseline cumulative hazards estimated by the naïve NIH-AARP, KW.S, and Post-KW versus the sample weighted NIHS is similar with the results shown in Table 5 on average (Web Figure 6).

# 6 DISCUSSION

Representative risk estimation for minority subgroups is challenging because, by definition, studies have small samples of minorities, who even worse, are often underrepresented in research cohorts. Hence, we proposed combining data sources, using data from nationally representative surveys to create pseudoweights for a cohort, then using a national registry to poststratify the pseudoweights. In contrast to "model-based" methods, our proposed post-KW.S PAR method relaxes the generalizability assumptions of naïve parameter estimates from cohort to the target population, although does require that the propensity model used to calculate pseudoweights is correctly specified. Poststratification is important to further reduce bias when the propensity model is misspecified (as shown in simulations and the data example) and to improve efficiency of pure risk estimation for minority subgroups, especially when cohort participation is informative. In simulations, as expected, the naïve Breslow estimate of risk, using only the cohort data, is biased, except when the cohort participation is non-informative (Scenario 1). The naïve PAR method for pure risks is also generally biased, even if cohort participation is non-informative, because the naïve attributable risks are biased. The post-KW.S PAR method obtains estimates of HRs and the pure risks with smallest MSE among the considered methods.

For developing an individualized model of all-cause mortality, NIH-AARP underestimated the HR for non-Hispanic Blacks and overestimated the HR for non-Hispanic others (who are mostly Asian-American), demonstrating that HR for minorities from non-representative studies cannot necessarily be generalized to the target population. Also, the original KW.S (without poststratification) overestimated the HR and pure risks for non-Hispanic others and underestimated the HR and pure risks for non-Hispanic Blacks. The NHIS survey, compared to CDC population summary statistics, overestimated pure risks for Hispanics and non-Hispanic others, probably due to small samples of deaths among minorities. Our findings suggest that HRs for minorities may not be generalized from a volunteer cohort to the general population, and that a survey can have inadequate sample sizes of minority subgroups for stable estimates. Our findings suggest to poststratify the pseudoweighted cohort to national registries to obtain more reliable risk estimates for minority subgroups.

Our key modeling assumption, correct specification of the propensity model, can be violated due to informative cohort participation, missing covariates, or incorrect functional form of the model. One can examine covariate distributions between the pseudoweighted cohort and weighted survey sample, but this is limited to common covariates in the two samples. In our example, KW.S did not obtain realistic HRs for minorities due to lack of fit of the propensity model for minorities or due to inadequate sample sizes of minorities in cohort or/and survey. HRs for minorities were not generalizable, which is a problem for "model-based" methods. A key value of poststratification is to ameliorate issues with the propensity model, which our simulations show is possible, but poststratification is limited to the variables available in the registry. Much research is being conducted on improving propensity models, such as using machine learning (Liu et al., 2021). More research on propensity model diagnostics is crucial for evaluating pseudoweighting-based methods in practice.

"Doubly robust" estimators are a promising approach to enhance robustness to possible misspecification of the propensity model (Chen et al., 2020, Lee et al., 2021), if the risk model were correctly specified. However, doubly robust estimators of pure risks and their corresponding analytical variances can be difficult to obtain. The complication comes from the complex data structure of combining the cohort and survey sample and the sampling designs used to collect the survey sample. In contrast, poststratification, which has been shown to improve efficiency and robustness to misspecification of the propensity model, is easy to implement for complicated point estimators, using existing software such as Poststratify() in survey package in R, without complex outcome model assumptions. Analytical variance estimation was derived using Taylor Linearization. Developing pure risk estimation that is doubly robust to the post-KW.S weights and the risk model is an area of future research.

In addition to developing more realistic propensity models and model diagnostics, other topics need further research. First, the proposed two-step weighting procedure could be extended to other epidemiologic study designs, such as case-cohort, nested case-control, and case-control studies, which are examples of multi-phase sampling designs (Smoot and Haneuse 2015). For example, the NCI Breast Cancer Risk Assessment Tool transports relative risks from a case-control study rather than a cohort. Second, time-dependent effects (e.g., lung cancer screening intervention services, Cheung et al., 2019) and time-varying risk factors (e.g., smoking status, Fisher and Lin, 1999) will be accommodated for pure risk estimation in future research. Also, our methods can be extended to estimating absolute risks that accounts for competing risks.

Figure 1 Log-Mortality rates over time by (a) age groups and (b) race/ethnicity groups in the CDC registry, 1997 NHIS, and NIH-AARP

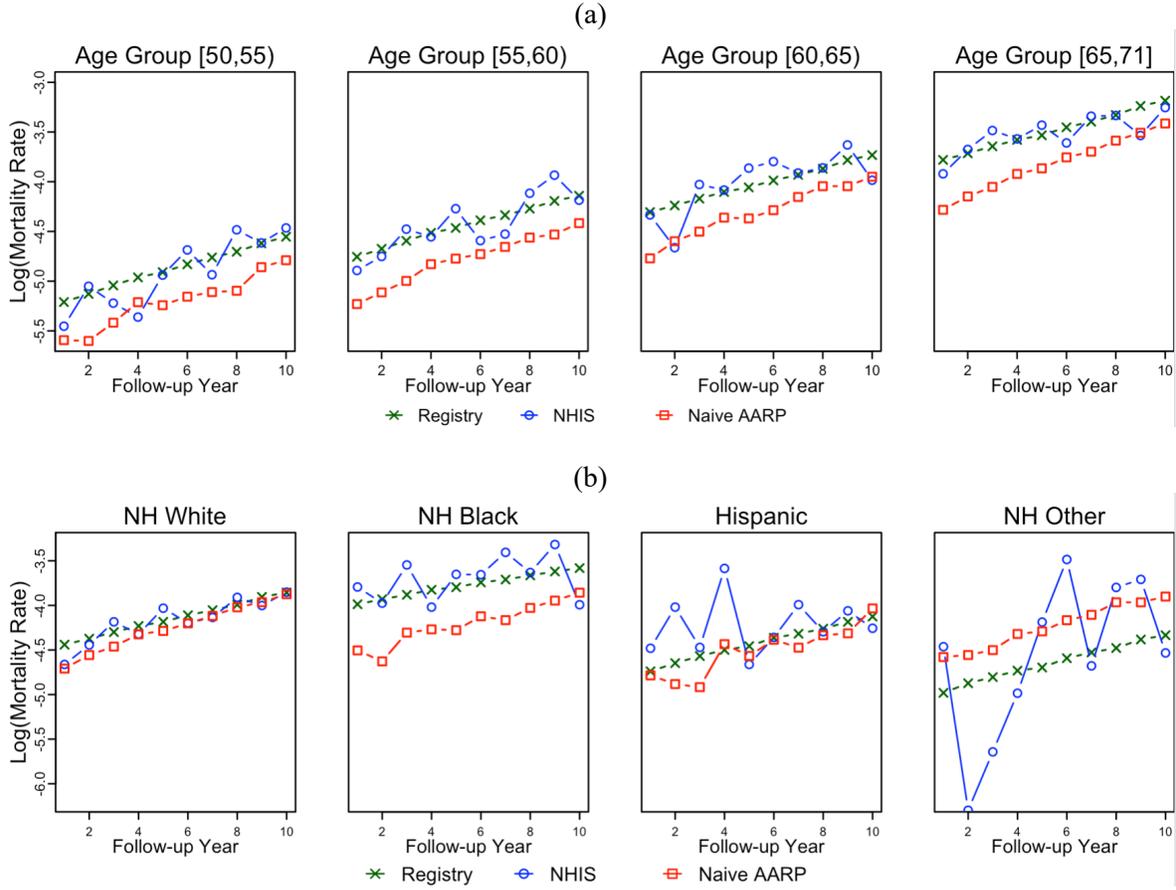

Table 1. Estimates of log-hazard-ratio ($\beta$) in Simulations

| Scenarios & Methods | Relative Bias[1] (%) | | | Empirical Variance[2] ($\times 10^{-4}$) | | | MSE[3] ($\times 10^{-4}$) | | |
|---|---|---|---|---|---|---|---|---|---|
| | $\beta_1$ | $\beta_2$ | $\beta_3$ | $\beta_1$ | $\beta_2$ | $\beta_3$ | $\beta_1$ | $\beta_2$ | $\beta_3$ |
| **Scenario 1** | Cohort participation rate $\propto \exp(0.1z_1 + 0.05z_2)$ | | | | | | | | |
| Naïve | -0.11 | -0.55 | -0.56 | 1.16 | 7.29 | 16.39 | 1.16 | 7.34 | 16.39 |
| Correctly specified propensity model: $\text{logit}(p) = \gamma_1 z_1 + \gamma_2 z_2$ | | | | | | | | | |
| KW.S | 0.18 | 0.09 | 0.35 | 1.61 | 8.63 | 18.93 | 1.61 | 8.63 | 18.93 |
| Post-KW.S | 0.22 | 0.17 | 0.29 | 1.51 | 4.49 | 18.96 | 1.51 | 4.50 | 18.96 |
| **Scenario 2** | Cohort participation rate $\propto \exp(0.1z_1 + 0.05z_2 + 0.3D)$ | | | | | | | | |
| Naïve | -4.37 | -4.31 | -4.75 | 0.86 | 5.71 | 12.53 | 2.05 | 8.78 | 12.97 |
| Correctly specified propensity model: $\text{logit}(p) = \gamma_1 z_1 + \gamma_2 z_2 + \gamma_d D$ | | | | | | | | | |
| KW.S | -3.15 | -2.69 | -3.21 | 1.24 | 7.01 | 15.12 | 1.86 | 8.22 | 15.32 |
| Post-KW.S | -0.69 | 1.25 | -0.69 | 1.23 | 4.08 | 15.7 | 1.26 | 4.34 | 15.71 |
| **Scenario 3** | Cohort participation rate $\propto \exp(0.1z_1 + 0.05z_2 + 0.3D - 0.1z_2 \cdot D)$ | | | | | | | | |
| Naïve | -2.84 | -21.94 | -3.74 | 0.92 | 6.23 | 13.27 | 1.42 | 85.89 | 13.54 |
| Correctly specified propensity model: $\text{logit}(p) = \gamma_1 z_1 + \gamma_2 z_2 + \gamma_d D - \gamma_{2,d} z_2 \cdot D$ | | | | | | | | | |
| KW.S | -0.35 | 0.55 | -0.38 | 1.36 | 13.87 | 16.23 | 1.37 | 13.92 | 16.23 |
| Post-KW.S | -0.34 | 0.30 | -0.38 | 1.30 | 5.58 | 16.16 | 1.31 | 5.60 | 16.16 |
| Misspecified propensity model: $\text{logit}(p) = \gamma_1 z_1 + \gamma_2 z_2$ | | | | | | | | | |
| KW.S | -1.95 | -22.27 | -2.41 | 1.33 | 7.56 | 15.80 | 1.57 | 89.57 | 15.91 |
| Post-KW.S | -0.13 | -8.66 | -0.66 | 1.30 | 4.31 | 16.31 | 1.30 | 16.70 | 16.32 |

[1] Relative bias $RB\% = B^{-1} \sum_{b=1}^{B} (\hat{\beta}^{(b)} - \beta_{FP}) / \beta_{FP} \times 100\%$, where $B = 10{,}000$ is the number of simulations, $\hat{\beta}^{(b)}$ $\hat{\beta}^{(b)}$ is the estimate of $\beta_{FP}$ obtained from the $b$-th simulated samples

[2] Empirical variance $V = (B-1)^{-1} \sum_{b=1}^{B} \{\hat{\beta}^{(b)} - B^{-1} \sum_{b=1}^{B} \hat{\beta}^{(b)}\}^2$

[3] Mean squared error $\text{MSE} = B^{-1} \sum_{b=1}^{B} \{\hat{\beta}^{(b)} - \beta_{FP}\}^2$

Table 2. Estimates of pure risks for three individuals with low, medium and high risks in simulations

| Method[1] | Relative Bias (%) | | | Variance ($\times 10^{-7}$) | | | Variance Ratio[2] | | | MSE ($\times 10^{-7}$) | | |
|---|---|---|---|---|---|---|---|---|---|---|---|---|
| | $r_{\text{low}}$ | $r_{\text{med}}$ | $r_{\text{high}}$ | $r_{\text{low}}$ | $r_{\text{med}}$ | $r_{\text{high}}$ | $r_{\text{low}}$ | $r_{\text{med}}$ | $r_{\text{high}}$ | $r_{\text{low}}$ | $r_{\text{med}}$ | $r_{\text{high}}$ |
| **Scenario 1** | Cohort participation rate $\propto \exp(0.1z_1 + 0.05z_2)$ | | | | | | | | | | | |
| Naïve (B) | -0.16 | -0.93 | -1.44 | 0.34 | 2.62 | 23.09 | 1.06 | 1.06 | 1.06 | 0.34 | 2.63 | 23.34 |
| Naïve (P) | -34.43 | -34.89 | -35.13 | 0.06 | 0.23 | 0.90 | 1.01 | 1.01 | 1.00 | 1.34 | 14.14 | 149.23 |
| Correctly specified propensity model: $\text{logit}(p) = \gamma_1 z_1 + \gamma_2 z_2$ | | | | | | | | | | | | |
| KW.S (B) | 0.04 | -0.29 | -0.32 | 0.42 | 3.21 | 28.39 | 1.01 | 1.00 | 1.00 | 0.42 | 3.21 | 28.40 |
| KW.S (P) | 0.47 | 0.22 | 0.28 | 0.14 | 0.45 | 1.58 | 1.03 | 0.97 | 0.88 | 0.14 | 0.45 | 1.59 |
| Post-KW.S (B) | -0.12 | -0.25 | -0.14 | 0.35 | 2.82 | 26.55 | 1.10 | 1.09 | 1.07 | 0.35 | 2.82 | 26.55 |
| Post-KW.S (P) | 0.25 | 0.13 | 0.27 | 0.12 | 0.39 | 1.60 | 1.01 | 0.93 | 0.87 | **0.12** | **0.39** | **1.61** |
| **Scenario 2** | Cohort participation rate $\propto \exp(0.1z_1 + 0.05z_2 + 0.3D)$ | | | | | | | | | | | |
| Naïve (B) | 36.89 | 29.53 | 22.74 | 0.51 | 3.53 | 28.19 | 1.04 | 1.04 | 1.05 | 1.98 | 13.50 | 90.32 |
| Naïve (P) | -30.24 | -33.92 | -37.23 | 0.05 | 0.18 | 0.69 | 1.02 | 1.01 | 0.98 | 1.04 | 13.33 | 167.23 |
| Correctly specified propensity model: $\text{logit}(p) = \gamma_1 z_1 + \gamma_2 z_2 + \gamma_d D$ | | | | | | | | | | | | |
| KW.S (B) | 0.49 | -0.16 | -0.55 | 0.35 | 2.64 | 23.47 | 1.02 | 1.00 | 0.98 | 0.35 | 2.64 | 23.51 |
| KW.S (P) | 1.15 | 0.55 | 0.21 | 0.12 | 0.37 | 1.38 | 1.05 | 1.00 | 0.87 | 0.12 | 0.38 | 1.38 |
| Post-KW.S (B) | 0.47 | 0.00 | -0.27 | 0.28 | 2.18 | 20.49 | 1.09 | 1.08 | 1.05 | 0.28 | 2.18 | 20.50 |
| Post-KW.S (P) | 0.92 | 0.45 | 0.20 | 0.10 | 0.32 | 1.38 | 1.07 | 1.04 | 1.00 | **0.10** | **0.32** | **1.39** |
| Misspecified propensity model: $\text{logit}(p) = \gamma_1 z_1 + \gamma_2 z_2$ | | | | | | | | | | | | |
| KW.S (P) | 8.24 | 4.12 | 0.42 | 0.12 | 0.37 | 1.31 | 1.01 | 0.95 | 0.92 | 0.20 | 0.56 | 1.33 |
| Post-KW.S (P) | 4.29 | 3.97 | 3.86 | 0.10 | 0.32 | 1.40 | 1.05 | 1.02 | 0.98 | 0.12 | 0.50 | 3.19 |
| **Scenario 3** | Cohort participation rate $\propto \exp(0.1z_1 + 0.05z_2 + 0.3D - 0.1z_2 \cdot D)$ | | | | | | | | | | | |
| Naïve (B) | 46.02 | 29.98 | 15.91 | 0.61 | 3.81 | 27.54 | 1.04 | 1.03 | 1.03 | 2.90 | 14.08 | 57.98 |
| Naïve (P) | -19.93 | -28.67 | -36.25 | 0.07 | 0.21 | 0.71 | 1.07 | 1.05 | 1.06 | 0.50 | 9.61 | 158.68 |
| Correctly specified propensity model: $\text{logit}(p) = \gamma_1 z_1 + \gamma_2 z_2 + \gamma_d D + \gamma_{2,d} D$ | | | | | | | | | | | | |
| KW.S (B) | 0.36 | -0.22 | -0.49 | 0.40 | 2.93 | 25.22 | 1.00 | 0.99 | 0.99 | 0.40 | 2.93 | 25.25 |
| KW.S (P) | 0.66 | 0.16 | -0.01 | 0.14 | 0.43 | 1.45 | 1.08 | 1.07 | 0.89 | 0.14 | 0.44 | 1.45 |
| Post-KW.S (B) | 0.59 | 0.12 | -0.12 | 0.30 | 2.42 | 22.68 | 1.09 | 1.07 | 1.04 | 0.31 | 2.42 | 22.68 |
| Post-KW.S (P) | 0.76 | 0.29 | 0.06 | 0.10 | 0.34 | 1.43 | 1.06 | 1.03 | 0.97 | **0.10** | **0.34** | **1.43** |
| Misspecified propensity model: $\text{logit}(p) = \gamma_1 z_1 + \gamma_2 z_2$ | | | | | | | | | | | | |
| KW.S (P) | 23.13 | 10.26 | -0.98 | 0.16 | 0.40 | 1.28 | 1.05 | 0.98 | 0.89 | 0.74 | 1.60 | 1.40 |
| Post-KW.S (P) | 10.50 | 6.15 | 2.21 | 0.12 | 0.35 | 1.40 | 1.04 | 1.02 | 0.95 | 0.24 | 0.78 | 1.99 |

[1](B) and (P) represents Breslow's and the PAR methods for cumulative baseline hazard estimation respectively.
[2]Variance ratio is the ratio of mean of Taylor variance estimates and variance of the risk estimates

Table 3. Estimates (95% confidence intervals) of hazard-ratios of all-cause mortality

|  | **NHIS 1997** | **Naïve AARP** | **KW.S** | **Post-KW.S** |
|---|---|---|---|---|
| **Age in years** (ref: 50 yrs) | 1.10 (1.08, 1.11) | 1.10 (1.10, 1.10) | 1.10 (1.09, 1.10) | 1.10 (1.09, 1.10) |
| **Sex** (ref: Male) | | | | |
| Female | 0.63 (0.55, 0.72) | 0.67 (0.66, 0.68) | 0.66 (0.64, 0.68) | 0.64 (0.63, 0.65) |
| **Race ethnicity** (ref: NH White) | | | | |
| NH Black | **1.34 (1.12, 1.61)** | **1.03 (0.99, 1.07)** | **1.01 (0.95, 1.07)** | **1.26 (1.21, 1.32)** |
| Hispanic | 0.99 (0.74, 1.27) | 0.82 (0.77, 0.87) | 0.77 (0.70, 0.84) | 0.83 (0.70, 0.86) |
| NH Other | **0.95 (0.68, 1.34)** | **1.01 (0.97, 1.06)** | **1.11 (1.04, 1.19)** | **0.65 (0.63, 0.67)** |
| **BMI** (ref: normal weight) | | | | |
| Underweight | 1.37 (0.85, 2.22) | 1.62 (1.52, 1.72) | 1.67 (1.32, 2.12) | 1.67 (1.48, 1.87) |
| Overweight | 0.80 (0.70, 0.91) | 0.89 (0.87, 0.90) | 0.84 (0.66, 1.07) | 0.84 (0.80, 0.87) |
| Obese | 0.84 (0.72, 0.99) | 1.01 (0.99, 1.03) | 0.94 (0.74, 1.20) | 0.94 (0.90, 0.99) |
| **Smoking** (ref: never smoking) | | | | |
| Former, quit ≥10 yrs | 1.06 (0.89, 1.26) | 1.31 (1.29, 1.34) | 1.32 (1.28, 1.37) | 1.29 (1.23, 1.35) |
| Former, quit <10 yrs | 1.97 (1.61, 2.40) | 2.04 (1.99, 2.09) | 2.02 (1.94, 2.09) | 1.95 (1.84, 2.06) |
| Current, <1pk/dy | 2.16 (1.83, 2.56) | 2.41 (2.35, 2.47) | 2.24 (2.15, 2.34) | 2.16 (2.02, 2.30) |
| Current, ≥1pk/dy | 2.70 (2.11, 3.44) | 3.00 (2.92, 3.09) | 2.82 (2.69, 2.95) | 2.70 (2.54, 2.87) |
| **Physical Activity** (ref: < 3 times/week) | | | | |
| >=3 times/week | 0.67 (0.55, 0.81) | 0.84 (0.82, 0.85) | 0.84 (0.82, 0.86) | 0.85 (0.82, 0.88) |
| **Health Status**[*] (ref: Excellent) | 1.58 (1.49, 1.67) | 1.66 (1.64, 1.67) | 1.74 (1.71, 1.76) | 1.71 (1.68, 1.75) |

[*]A five-level trend variable with scores (Excellent=0, very good=1, good=2, fair=3, poor=4)

Table 4. Ratio of (Weighted) mean of 10-year pure risk estimates (expected) and CDC mortality rates (observed) by demographic variables available in CDC data.

| | Poststrata | NHIS PAR | AARP Breslow | AARP PAR | KW.S PAR | Post-KW.S PAR |
|---|---|---|---|---|---|---|
| **Overall** | | 0.98 | 0.88 | 0.97 | 0.96 | 0.97 |
| **Age** | 50-54 | 0.94 | 0.73 | 0.82 | 0.94 | 0.98 |
| | 55-59 | 0.95 | 0.72 | 0.80 | 0.92 | 0.97 |
| | 60-64 | 0.98 | 0.75 | 0.83 | 0.95 | 0.99 |
| | 65-71 | 0.95 | 0.72 | 0.80 | 0.94 | 0.96 |
| **Sex** | Male | 0.98 | 0.84 | 0.93 | 0.98 | 0.97 |
| | Female | 0.97 | 0.85 | 0.94 | 0.95 | 0.97 |
| **Race Ethnicity** | NH White | 0.94 | 0.89 | 0.98 | 0.98 | 0.98 |
| | NH Black | 1.03 | 0.65 | 0.72 | 0.72 | 0.95 |
| | Hispanic | 1.22 | 0.90 | 1.00 | 0.93 | 0.96 |
| | NH Other | 1.35 | 1.43 | 1.58 | 1.75 | 0.97 |

Table 5. Estimates (standard errors $\times 10^{-2}$) of pure risk of mortality in ten years for three minority individuals with low, medium, and risks[1] from AARP and NHIS

| Risk Level (Race) | NHIS PAR | AARP Breslow | AARP PAR | KW.S PAR | Post-KW.S PAR |
| --- | --- | --- | --- | --- | --- |
| Low (Hispanic) | 0.050 (1.471) | 0.039 (0.177) | 0.044 (0.197) | 0.045 (0.311) | 0.051 (0.367) |
| Medium (NH-Black) | 0.190 (3.888) | 0.241 (0.553) | 0.267 (0.595) | 0.222 (1.008) | 0.134 (0.690) |
| High (NH-Other) | 0.874 (4.073) | 0.766 (0.825) | 0.804 (0.771) | 0.810 (1.180) | 0.873 (0.977) |

[1]The covariate values of the three individuals are (1) Low-risk: 50-year-old, male, Hispanic, underweighted, former smoker (quit yrs<10), ≥3 times/week physical activities, excellent self-reported health status; (2) Medium-risk: 60-year-old, male, Non-Hispanic other, obese, current smoker (<1 pack/day), ≥3 times/week physical activities, good self-reported health status; and (3) High-risk: 70-year-old, female, Non-Hispanic Black, overweight, current smoker (>1pack/day), <3 times/week physical activities, fair self-reported health status.